\documentclass[10pt]{aastex}
\usepackage{bm}
\usepackage{lscape}
\usepackage[dvips]{epsfig}
\usepackage{amsmath}
\usepackage{amssymb}
\usepackage{amsthm}
\usepackage{array}
\usepackage{multirow}
\usepackage{graphicx}
\usepackage{amsfonts}
\usepackage{eucal}
\newcommand{\ms}{$M_{\odot}$}
\newcommand{\msb}{$M_{\odot}$~}
\newcommand{\ct}{$^{13}$C}
\newcommand{\ctb}{$^{13}$C~}
\newcommand{\ctanb}{$^{13}$C($\alpha$,n)$^{16}$O~}

\newcommand{\neanb}{$^{22}$Ne($\alpha$,n)$^{25}$Mg~}

{}
\begin{document}

\title{$s$-Processing in AGB Stars Revisited. I. Does the Main Component Constrain the Neutron Source in the $^{13}$C-Pocket?}

\author{O. Trippella, M. Busso}
\affil{Department of Physics, University of Perugia, and INFN, Section of Perugia,
via A. Pascoli, 06123 Perugia, Italy; oscar.trippella@fisica.unipg.it; maurizio.busso@fisica.unipg.it}

\author{E. Maiorca}
\affil{INAF, Observatory of Arcetri, Viale E. Fermi 5, 50125 Florence, Italy and INFN, Section of Perugia, via A. Pascoli, 06123 Perugia, Italy}

\author{F. K\"appeler}
\affil{Karlsruhe Institute of Technology, Campus North, Institute of Nuclear Physics, P.O. Box 3640, 76021 Karlsruhe, Germany}

\author{S. Palmerini}
\affil{INFN, Laboratori Nazionali del Sud, via Santa Sofia 62, 95125 Catania, Italy}

\begin{abstract}
Slow neutron captures at $A \gtrsim 85$ are mainly guaranteed by the reaction \ctanb in AGB stars, requiring proton injections from the envelope. These were so far assumed to involve a small mass ($\lesssim 10^{-3}$ \ms), but models with rotation suggest that in such tiny layers excessive $^{14}$N hampers $s$-processing. Furthermore, $s$-element abundances in Galaxies require \ct-rich layers substantially extended in mass ($\gtrsim 4 \times 10^{-3}$ \ms). We therefore present new calculations aiming at clarifying those issues and at understanding if the solar composition helps to constrain the \ctb ``pocket'' extension.  We show: i) that mixing ``from bottom to top'' (like in magnetic buoyancy or other forced mechanisms) can form a \ctb reservoir substantially larger than assumed so far, covering most of the He-rich layers; ii) that stellar models at a fixed metallicity, based on this idea reproduce the main $s$-component as accurately as before; iii) that they make nuclear contributions from unknown nucleosynthesis processes ($LEPP$) unnecessary, against common assumptions. These models also avoid problems of mixing at the envelope border and fulfil requirements from C-star luminosities. They yield a large production of nuclei below $A = 100$, so that $^{86,\;87}$Sr may be fully synthesized by AGB stars, while $^{88}$Sr, $^{89}$Y and $^{94}$Zr are contributed more efficiently than before. We finally suggest tests suitable to say a final word on the extension of the \ctb pocket.
\end{abstract}


\section{The state of the art.}
A few years ago it was believed that the ``$s$ process'' was essentially clarified. It had been understood many years before that the distribution of $s$-elements in the Solar System required multi-modal neutron exposures \citep{war78}, generated in different astrophysical sites.
A {\it main} component (accounting for nuclei from Sr to Pb) had been attributed to stars climbing for the second time along the red giant branch in a phase called Asymptotic Giant Branch, or AGB \citep{ibe75,gal88}, while a {\it weak} component (explaining nuclei up to the $N$ = 50 magic neutron number) had been ascribed to massive stars, during core-He and shell-C burning \citep{rai93}. Recent work on the weak component \citep[in particular by][] {the07,pig10} showed that this process is complex and strongly dependent on reaction rate uncertainties affecting $\alpha$-captures (especially the $^{12}$C+$\alpha$ reaction), the $^{12}$C+$^{12}$C fusion and neutron captures on nuclei immediately beyond iron. Hence, final results for the weak $s$-component must necessarily wait for new measurements and might, in the mean time, profit of any improvement in the constraints from the main component.  

A third, {\it strong} component, originally devised to account for 50\% of the $^{208}$Pb concentration \citep{cla67} was recognized to be unnecessary \citep{gal98}, as its role could actually be played by AGB stars at low metallicity, where the scarcity of iron nuclei implies 
that the number of neutrons per iron seed easily becomes sufficiently large to feed Pb. 
This passage from the main to the strong component in AGB stars is gradual and depends on the efficiency of the neutron release, so that heavy nuclei become progressively more enriched for decreasing metal content \citep{tra01,bus01}.

A series of works presented in the last twenty years also clarified that the main component produced by AGB stars is due to the activation of the \ctanb source and (more marginally) of the \neanb source \citep{kae90}. Evidence on the minor role of the \neanb reaction was provided by the abundances of neutron-rich, stable isotopes of heavy elements in the Solar System (especially $^{86}$Kr and $^{96}$Zr). Indeed, when the neutron release from the \neanb reaction was allowed to increase (a fact that requires relatively high temperatures, i.e. $T \geqslant 3.5 \times 10^{8}$ K), excessive production of these nuclei invariably resulted \citep{kae90,arl99}. Another constraint pointing to the same direction came from the Rb/Zr ratios in carbon stars \citep{abia01,abia02}.
This suggested that the parent stars had to be of relatively low mass, i.e. $M \lesssim 3$ \ms, because for this mass range the typical temperature in the pulses barely reaches $T = 3 \times 10^8$ K. In these stars, thermonuclear combustion in the He-burning shell is activated in short, explosive bursts (\textit{thermal pulses}), during which an intermediate convective zone forms in the He-rich layers; these bursts are separated by long intervals (several 10$^4$ yr), called \textit{interpulse} periods, where the \ctanb source releases its neutrons and produces $s$-elements \citep[see reviews in][]{bus99,kae11}. 

Even now, the above scenario is not exempt from weak points. In particular, the activation of the \ctanb neutron source requires unclear partial mixing mechanisms injecting protons into the He-rich layers during the downward envelope expansion called \textit{third dredge-up} (TDU), occurring after advanced thermal pulses. The lack of knowledge on these phenomena forced researchers to parameterize the amount of \ctb available in a relatively free way.

The scenario outlined above was subsequently put in question by chemical evolution models of $s$-elements in the Galaxy by \citet{tra04}. These authors could not obtain, from the integrated Galactic production, the good fit to the main $s$-process component previously envisaged on the basis of a single AGB model for a Low Mass Star (LMS) of a suitable metallicity. The Galactic production reconstructed using nuclear yields from AGB models with a rather small ($\simeq 10^{-3}$ \ms) \ct-pocket turned out to be insufficient to account for the solar abundances of n-rich elements with atomic mass numbers between 86 and 130. In a recent work \citet{bis14} confirm these results adopting a \ct-pocket extension in the range from 2.5$\times$10$^{-4}$ to 2$\times$10$^{-3}$ \ms.

Very recently, observations of $s$-process abundances were obtained for a large sample of Galactic open clusters, first for Ba \citep{dor09}, then also for Y, Zr, La and Ce \citep{mai11}. As the age of an open cluster can be determined accurately, the above authors could trace, for the first time, the evolution of $s$-elements in the Galactic disk directly as a function of time. This allowed them to observe an unexpected growth of $s$-element abundances in young stars of the Galaxy. Subsequent works on open clusters by \citet{yon12,jac13,mis13}, performed with different analysis methods, confirmed that such an increase is in general real (although for elements like Zr and La further investigations are needed).  Another piece of evidence in the same direction came from a complementary study by \citet{mcw13}, who found $s$-element abundances increasing with time in the Sagittarius Dwarf Galaxy. They showed that the traditional slow neutron-capture nucleosynthesis scenario fails to reproduce the observed trend, which instead requires a larger neutron inventory in AGB stars. This confirms a recent proposal by \citet{mai12}; they suggested that the \ctb reservoir formed at TDU be considerably larger than previously adopted, at least in AGB stars of very low mass ($M \lesssim 1.5$ \ms). With this assumption and a chemical evolution model for the Galaxy, they could reproduce very well the abundances observed in open clusters.

Another source of doubts on the traditional way of modeling the $s$-process emerged recently, this time from the theoretical side. In a paper by \citet{pie13} it was shown that, by including rotation in stellar modeling, any partial mixing at the convective border (like those previously invoked for forming a {\it small} \ctb pocket) become affected by the Goldreich$-$Schubert$-$Fricke instability, forcing a more complete mixing. Hence, at hydrogen shell re-ignition, any layer affected by the penetration of envelope material becomes dominated not by $^{13}$C, but by $^{14}$N, which is an efficient neutron poison and would strongly reduce the $s$-process efficiency. 
This confirms previous suggestions by \cite{lan9,her3,sies4} and
indicates that all previous attempts at modelling the \ctb pocket formation through a small ($\lesssim 10^{-3}$ \ms) exponential penetration of protons below the envelope border \citep{cri09,cri11} would be no longer acceptable. Notice that in any case, this proton penetration had already the critical property of being dependent on the TDU phenomenon, being a downward partial extension of it. This induced in any case a limit on the \ctb pocket formation, as it could occur only after pulses followed by TDU, i.e. in rather advanced stages of the AGB.

In this paper we want to re-analyze the rather confused situation, which emerged from the above discoveries, by ascertaining:
i) if the hypothesis of a \ctb pocket substantially larger than imagined in the past years is compatible with a reproduction of the $whole$ distribution of $s$-elements in the solar main component \citep[as the][paper only verified this point for Y, Zr, Ba, La, Ce]{mai12}.
ii) if one can imagine forms of deep mixing alternative to the partial extension of the envelope, in order to suggest ways for putting the formation of the \ctb pocket on safer grounds; and iii) if an analysis of the solar-system main $s$-process component can constrain the extension of the layers partially polluted by protons at TDU, where \ctb is expected to form.

In section 2 we present some simple ideas aimed at setting the stage for a physical modeling of the \ctb pocket. In section 3 we discuss recent improvements in the nuclear inputs, both for neutron-capture cross sections and for the rates of the neutron-producing reactions. In section 4 we describe our computations for $s$ processing in low-mass AGB models at suitably-chosen  metallicities, supposed to represent an average of the $s$-processing efficiency over the evolution of the Galaxy. In this way we adopt, for the sake of comparison, two widely different extensions for the \ctb pocket (in the range of those so far proposed by different authors). This is aimed at understanding how these results compare with the solar distribution, an issue that is discussed in section 5. Finally, in section 6 we discuss the implications of the results found and we suggest some observational and theoretical tests that should help saying a final word on the \ctb formation in AGB stars.

\section{A possible way out for the \ctb pocket.}

We shall try here to argue in favor of the existence of a \ctb reservoir in He-rich layers, despite the doubts recently advanced on the poorly known physics of the convective border at TDU.
One can rely for this purpose on very old (but seminal and too often forgotten) discussions of gas and plasma physics in stars, by S. Chapman, T.G. Cowling and E.N. Parker. In particular, \citet{par58}
showed that, when in a stellar layer a suitable ``engine'' exists (i.e. a local extra-source of heating, which adds to the general energy production sited deep in the star), then the mechanical behavior of that
region will depend on how its temperature drops with the radius. Parker noticed that sufficiently far from the input energy source, considerations on the heat flow 
require that:
$$
T(r) = T_0 \left( {\dfrac{r_0}{r}} \right)^{1/(n+1)} \eqno(1)
$$
where $r_0$ is the radius within which all the extra-energy is provided \citep[see equation 3 in][]{par58}. He also
showed, from heat exchange considerations (equation 1 in the quoted paper), that $n$ must be positive.
The mentioned paper considered a gas of pure hydrogen, but its results have a general application,
and $n$ must be positive also for gases made by admixtures of heavier ions. This is shown in detail in
\citet{cha51}, especially in chapter 18 (integrated by equation 12.1.I of the same book). This generalization
is crucial for us, in view of the fact that, below TDU, the material we deal with is mainly made of $^{4}$He
and $^{12}$C. As \citet{par58} showed, equation (1) with $n \gtrsim$ 0 is not compatible with a hydrostatic
solution, so that the material is in a state of {\it natural} expansion.
\citep[By applying these ideas to the hot solar corona, Parker then gave the first quantitative physical basis to the existence of the solar wind:][]{par60}.

We can now notice that the temperature condition for Parker's solution, i.e. $T \propto r^{-\beta}$, with $\beta \lesssim 1$, holds also in the AGB layers of our interest, during TDU.
This is shown in  Figure 1, where we plotted the behavior of pressure, density and temperature for the mentioned zones of a 1.5 \msb model by \citet{str03}, during the fourth TDU (the trends plotted are actually typical of all AGB stars in the mass range from 1 to about 3 \ms). As is clear from the third panel, the temperature decreases with radius less rapidly than $r^{-1}$ (the actual exponent varying between $-0.93$ and $-0.2$).

In the above conditions, {\it if there is an extra source of  energy acting at the base}, the overlying zones can be put in an expanding motion with respect to the environment, due to the Parker's mechanism.
(Notice that the He-rich layers are {\it already} expanding as a consequence of the nuclear energy input introduced shortly before TDU by the occurrence of a thermal pulse).

\begin{figure}
\centering
\includegraphics[scale=0.45]{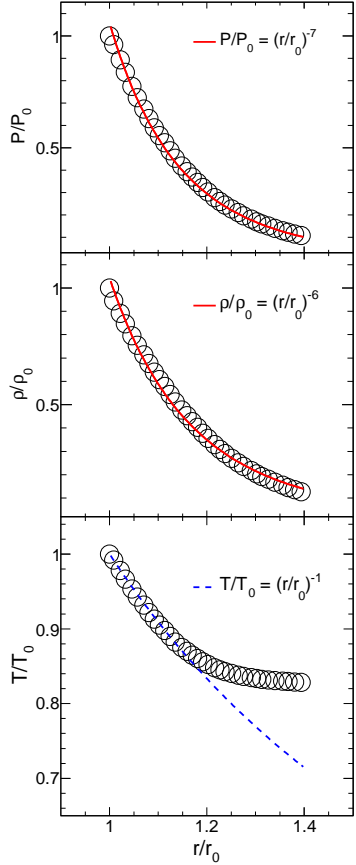}
\caption{The pressure (${\rm P}$), density (${\rm \rho}$) and temperature (${\rm T}$) distributions as a function of radius in the \textit{intershell} region of a star with 1.5 \msb and solar metallicity. The physical quantities are normalized to their values at the bottom of the He-rich region. The top and the middle panel jointly show that the stellar structure is a polytrope (${\rm P} \propto {\rm \rho}^{7/6}$). The red and full lines are the best fits to the ${\rm P}$ and ${\rm \rho}$ trends found in the stellar model, respectively. In the lowest panel, the blue-dashed line is a guide to the eye. The temperature decreases less rapidly than $1/{\rm r}$ over the whole layer of interest,
thus satisfying the conditions required by Parker. The formulae describing the fitting lines are
shown at the upper right corner of each panel. All quantities are divided by the corresponding
values (with subscript index ``${}_{{\rm 0}}$'') at the layer just above the C-O core.\newline
(A color version of this figure is available in the online journal.)}\label{one}
\end{figure}

Let's now consider a rotating AGB star, with a rigid-body inner degenerate core and an intermediate,
differentially-rotating layer below a convective envelope. This structure can power a magnetic dynamo
\citep{nor08}, suitable to induce the buoyancy of magnetized domains that reach the envelope \citep{bus07},
as it actually occurs in the Sun \citep{par84}. Motivations in favor of magnetic buoyancy as
a source of non-convective mixing have been presented elsewhere for H-rich layers \citep{den09,dm11} and
will not be repeated here.

A further energy input is then provided by the formation of toroidal magnetic fields  near the border of the rigid-body core by a dynamo mechanism \citep{bus07,nor08}. In the magnetized regions  thus formed (which will in general occupy a fraction $f_r$ of the total mass of a stellar layer of radius $r$) the extra-source of energy will be the magnetic energy density, i.e. the magnetic pressure, equal (in cgs units) to $B_r^2/8\pi$ (where $B_r$ is the magnetic field in the magnetized zones at the layer $r$).

Several papers then described the {\it relative} buoyancy of magnetized structures (usually,
magnetic flux tubes) with respect to an underlying, non-magnetized gas \citep[see e.g.][]{par74}.
In our case buoyant flux tubes crossing the convective envelope border thanks to the extra-energy
provided by magnetic pressure would behave somehow similarly to the solar wind, because the conditions
set by \citet{par58} and \citet{cha51} are satisfied. One can then express the rate of buoyancy for the
magnetized mass crossing a stellar surface at radius $r$ as:
$$
\dot M =  4 \pi r^2 \rho v_r f_r\eqno(2)
$$
Conservation of the rising mass would yield $v_r f_r$ = const. Conservation of mass across the convective
border would instead guarantee that a downward flux of envelope material occurs, with 71\% in mass of protons
(for a solar composition). This will not be due to some dubious, spontaneous smoothing of the convective
velocity profile, but to a {\it forced} process, the  ``engine'' residing down near the degenerate
core, rooted in a magnetic dynamo. This is promising: measurements of $B$ (from which the
rising velocity will depend) might in a near future fix the mass circulation, hence the amount of protons
penetrating downward. In other words, the \ctb pocket might be fixed consequently.


Ours is not a demonstration: it cannot be, without a quantitative, detailed magneto-hydrodynamic (MHD) modeling. However, we believe that our suggestion may deserve further scrutiny with detailed models, as MHD might provide the required forcing term, suitable to push 
down from the envelope into the radiative layers, for mass conservation, the protons we need for forming the \ctb pocket.

One has to notice that the material pushed down from the envelope must move against a pressure and density gradient. Quite generally, this will lead to a decrease in the penetrated mass with distance, as the environment will provide a gradual extinction for the flow. If this can be expressed via a constant extinction coefficient $\alpha$  (depending on viscosity, on the degree of thermalization, etc...) then the problem can be treated similarly to any transport process. Calling $dM$ the mass that can travel downwards a length $dr$ inside the He-rich layers, one has:
$$
\frac{dM}{M} = -\alpha dr \eqno(3)
$$
If the forced penetration reaches down to a depth $r$ and if the base of the envelope is sited at the radius $r_E$, then for any injection of a mass $M_0$ at $r_E$ integration of equation (3) yields:
$$
M(r)=M_0e^{-\alpha(r-r_E)}
\eqno(4)
$$
This assumes $\alpha$ to be approximated by a constant. It is however plausible that the viscosity and the pressure gradient grow with the distance from the envelope; 
then the corresponding $\alpha(r)$ would also grow and the mass profile of the penetrating material would be steeper than a simple exponential. For our merely illustrative purposes we can stick to the simple case of equation (4). Notice that the MHD hypothesis does not enter into equation (4). That idea is promising, but it serves us only to make plausible that a \ctb pocket can in fact be produced in Galactic disk AGB stars, to compensate the problems now emerging in the traditional approach \citep[see][in particular table 2 and section 5]{pie13}, by a forced mechanism, and can for this reason be larger than so far adopted. From this point of view, other mechanisms might exist to serve the same purpose.  Perhaps another promising example would be the wave-like form of mixing suggested by \citet{dt03}. The important point is that, for explaining a large \ctb reservoir, one should look for a $forced$ process, not a $free$ one.

For the sake of comparison with previous works, we shall assume, for the moment, that the extension of the \ctb pocket,ensuing from a forced mechanism of the kind qualitatively motivated above, be either similar to the one adopted
by \citet{bus01}, or much larger, as suggested by \citet{mai12}. These last authors derived, from chemical evolution models for the Galaxy, the requirement  for a pocket extension in mass from 3 to 8 times larger than before (depending on which of the older models is assumed as a comparison). We shall adopt here, as an example, a pocket of 6$\times 10^{-3}$ \ms. Notice that this means polluting with protons almost the entire extension of the He-rich layers, as required by physical mechanisms driven by buoyancy from the levels where a dynamo is established.

For the pocket used in traditional models we chose an extension aimed at representing a sort of average among the many cases previously considered. Our choice is larger than in \citet{tra04}; it is similar to that of Case A in the recent paper by \citet{bis14} and is only a factor-of-two smaller than for their maximum extension. Concerning the total mass of \ctb available for burning, in the smaller pocket adopted by us for comparison this is about $3.45\times10^{-6}$ \ms. 

Figure 2, instead, shows the abundances in the more extended pocket of our second hypothesis, after H-shell reignition. The pocket itself was obtained by introducing mass from the envelope with the parameter $\alpha$ of equation (3) set to reproduce the proton penetration suggested by \citet{mai12}; it extends by about 6$\times 10^{-3}$ \ms. It contains 4.2$\times 10^{-5}$ \msb of \ct, almost entirely confined in the first 4$\times10^{-3}$ \ms. The extension is more than a factor of 3 larger than the largest case discussed by \citet{bis14}. Notice that these authors find the pocket extension to be irrelevant for the chemical evolution of the Galaxy. However the maximum abundance of \ctb they considered does not exceed 1.1$\times$10$^{-5}$ \ms, a value that is a factor of four smaller than our case of Figure 2. It is therefore not surprising that \citet{bis14} did not find any significant difference as compared to older models by \citet{tra04}.

In our approach, the extension of the pocket is a direct consequence of the physical structure of the star and the abundances in it are constrained by the rates for proton captures. The forced mechanism ``from bottom to top'' we consider will cover most of the He-rich layers and will therefore fix, for each stellar mass and metallicity, essentially an unique value for the number of neutrons produced, We notice that this fact avoids a free parameterization, often introduced by allowing each stellar model at each mass and metallicty to host a range of different $s$-process efficiencies \citep{tra04}.


As is clear from Figure 2, the upper part of the reservoir, initially containing more protons, gets enriched in the neutron-poison  $^{14}$N, while in the inner part $^{13}$C dominates. This second region is where most $s$-elements are produced, while the upper part is especially important for the complex nucleosynthesis network starting from $^{14}$N, feeding $^{15}$N, $^{19}$F and $^{23}$Na \citep{cri11}. The contribution of this upper layer
to the $s$-process is smaller, as only about 30\% of the neutrons are saved to be captured by heavy seeds. Nevertheless, this zone is very important for the solar distribution: as an example, when the \ct-rich layer is suitable to feed nuclei near the $N = 82$ peak (e.g Ba), this zone rich in $^{14}$N will mainly feed the lighter $s$-elements, like Sr and Zr: considering it properly and is therefore crucial for the synthesis of nuclei below A =100 or so.

\begin{figure}
\centering
\includegraphics[scale=0.6]{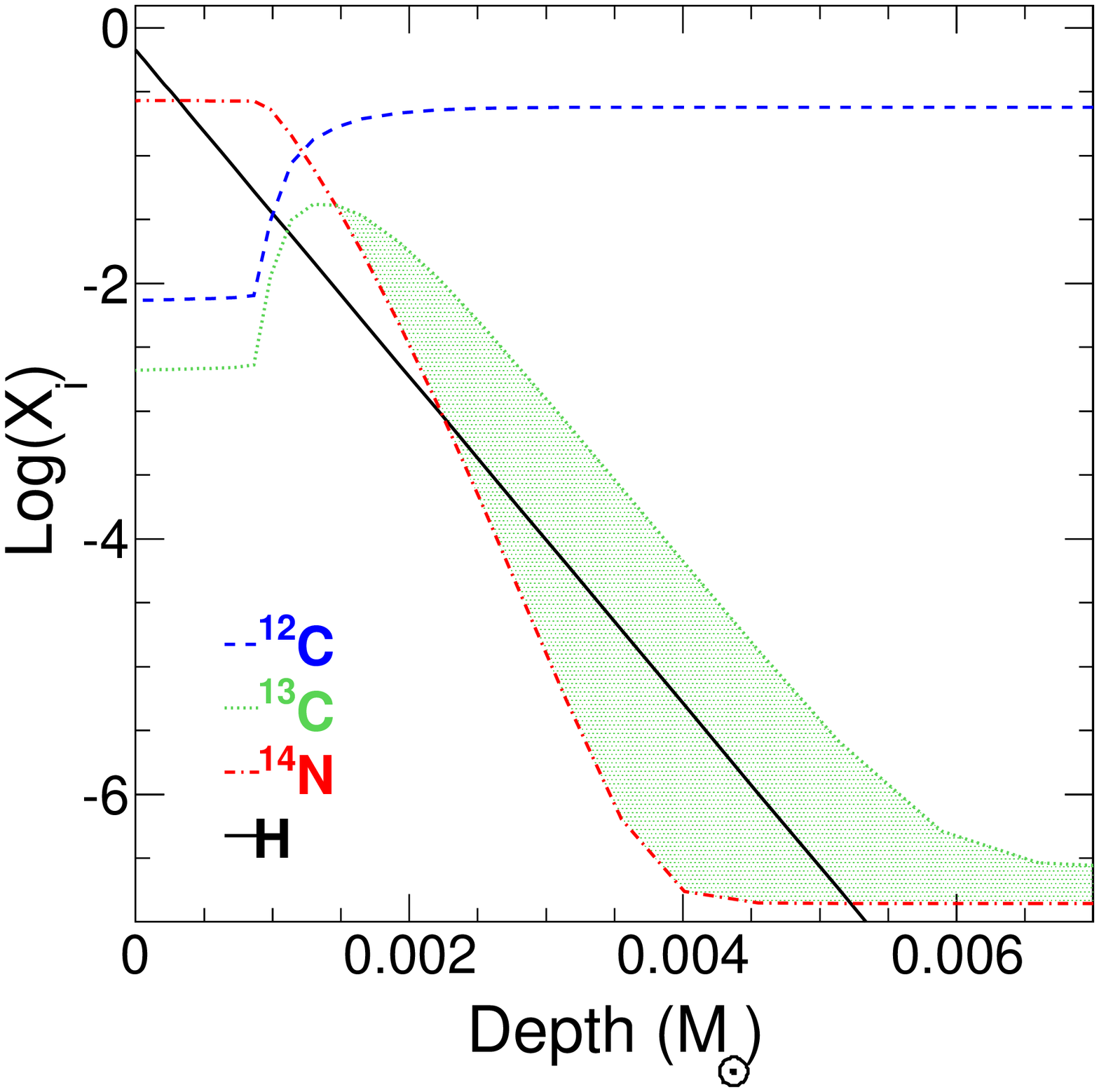}
\caption{The extended $^{13}$C pocket formed according to the model described in the text. The mass extension is $6 \times 10^{-3}$ \msb. It is more extended by a factor from 3 to 8 with respect to cases common in the previous decades \citet{gal98,tra04}. From the outside towards the inside of the star (or equivalently from left to right in the figure) we obtain first a region where $^{14}$N (red dot-dashed line) dominates, then a second layer (green-shaded in the figure) where the $^{13}$C abundance prevails. \newline
(A color version of this figure is available in the online journal.) } \label{two}
\end{figure}

\section{On the nuclear inputs adopted.}

As mentioned, the ${}^{13}{\rm C}(\alpha,{\rm n}){}^{16}{\rm O}$ reaction is the dominant source of neutrons for the production of the $s$-process main component in low-mass stars. There is a vast literature on this reaction, as it presents challenging experimental problems. The main one concerns the presence of a subthreshold resonance at $-3$ keV in the center-of-mass system corresponding to the excited 1/2$^+$ state of $^{17}$O. Direct measurements have been so far possible down to 270 keV; below this energy only theoretical extrapolations are available. The Gamow window lays in the range  140 $-$ 230 keV, where the astrophysical $S(E)$-factor is dominated by the subthreshold resonance. The effect of this resonance was investigated in  several experiments using indirect techniques. The spectroscopic factor was determined by \citet{kub03} and then revisioned by \citet{kee03}, while three experiments by \citet{pel08}, \citet{joh06} and \citet{guo12} extracted or calculated the Asymptotic Normalization Coefficient (ANC) from the same resonance. The resulting reaction rate showed, however, large differences. New direct measurements of the ${}^{13}{\rm C}(\alpha,{\rm n}){}^{16}{\rm O}$ cross section were performed by \citet{hei08b} to improve the data at higher energy ($E_{c.m.}$ = 320 $-$ 700 keV). These authors measured also the double differential cross section for elastic scattering, ${}^{13}{\rm C}(\alpha,\alpha){}^{13}{\rm C}$, at 28 angles between $E_{lab}$ = 2.6 $-$ 6.2 MeV to constrain possible contributions from background resonances. By normalizing previous data \citep{ang99} to their results, \citet{hei08b} performed a comprehensive $R$-matrix fit to reduce the uncertainties in the extrapolation of $S(E)$ down to very low energies.
Recently, a new indirect measurement \citep{lac12,lac13} with the Trojan Horse Method \citep[hereafter THM, see][for a review]{spi11} was derived using the ${}^{13}{\rm C}({}^{6}{\rm Li},{\rm n}{}^{16}{\rm O}){\rm d}$ reaction at energies above the Coulomb barrier for extracting the two-body astrophysical factor. The unambiguous observation of the subthreshold resonance at $-3$ keV allowed the extraction of the ANC for the first time in a Trojan Horse experiment \citep{lac12}, thus joining the two indirect techniques. The OES (\textit{off-energy-shell}) $R$-Matrix fit to $S(E)$ was performed adopting the same procedure described in \citet{hei08b}, using four resonances above the Coulomb barrier for the global fit. In Figure 3 we compare the new THM reaction rate to other data currently used in nucleosynthesis calculations. All rates are divided by the THM recommended value. The new rate is higher than those by \citet{dro93}, \citet{cf88}, \citet{hei08b}, so that the $^{13}$C burning time-scale is slightly shorter than previously assumed. On the contrary, the new rate is lower than in the NACRE compilation. In this work we adopt the analytical expression provided by \citet{lac13}, because of its small uncertainty. This is actually the first time that this new rate is used in nucleosynthesis calculations.

\begin{figure}
\centering
\includegraphics[scale=0.6]{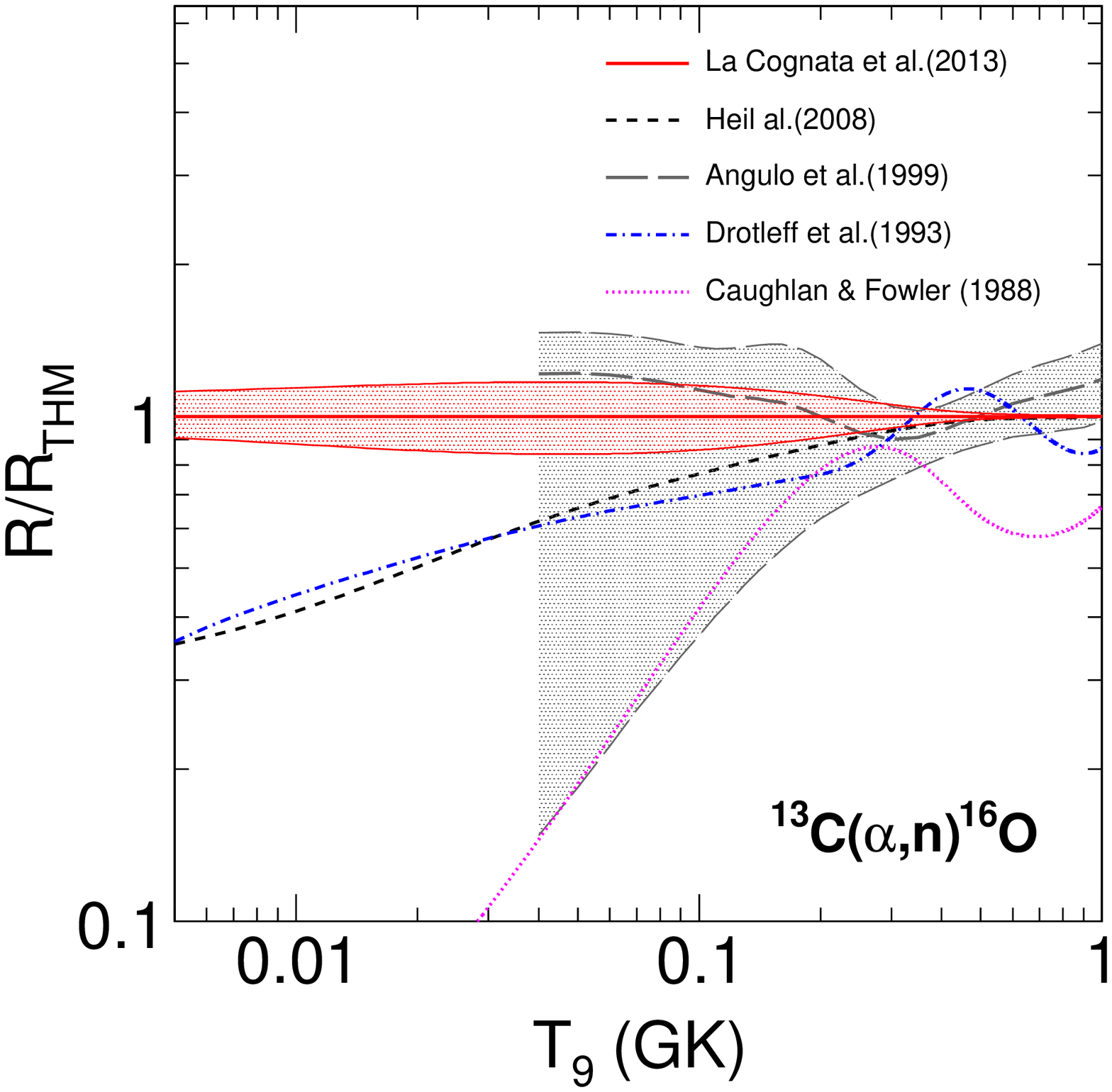}
\caption{Ratios between some of the most commonly used estimates for the ${}^{13}{\rm C}(\alpha,{\rm n}){}^{16}{\rm O}$ reaction rates and the one by \citet{lac13}, adopted in our $s$-process nucleosynthesis calculation. The continuous red line represents the data by \citet{lac13} with the corresponding uncertainties shown by the shaded red area, while the long-dashed grey region refers to NACRE \citep{ang99}. The reaction rates by \citet{hei08b}, \citet{dro93} and \citet{cf88} are represented by short-dashed black, dot-dashed blue, and dotted magenta lines, respectively.\newline
(A color version of this figure is available in the online journal.)} \label{three}
\end{figure}

The second relevant neutron source for AGB stars is the ${}^{22}{\rm Ne}(\alpha,{\rm n}){}^{25}{\rm Mg}$ reaction. At temperatures typical of He burning in a low mass star (around 2.8 $-$ 3 $\times$ 10$^8$ K at most) it is only marginally activated, but produces an additional burst of neutrons, during convective thermal pulses, which helps in fixing the abundances near reaction branchings \citep{kae90,arl99}. In recent years the main uncertainty in the ${}^{22}{\rm Ne}(\alpha,{\rm n}){}^{25}{\rm Mg}$ cross section has been related to the resonance at 633 keV, which might or not give some contributions at low energy. In this framework, \citet{jae01} performed measurements using a target enriched to 99.9\% in ${}^{22}{\rm Ne}$ and a $4\pi$ neutron detector to measure the excitation function from 570 to 1550 keV. In this way, the parameters of the resonances were extracted and an analytical formula for the reaction rate was provided. Its values are lower, at the relevant energies, than in \citet{kae94} and in NACRE \citep{ang99}, up to a factor of two. In this work we adopted the recommended value of \citet{lon12}, which was obtained in an updated Monte Carlo analysis including all previous data, thus superseding earlier results presented in \citet{ili10}. The ratio of this choice (${\rm R}_{\rm L}$) is compared with other data in panel a) of Figure 4. In the energy region relevant for low-mass AGB stars the rate of \citet{lon12} is about 25\% lower than the one suggested by \citet{kae94}, due to the lack of any effect from the crucial resonance at 633 keV. However, one has to notice that all measurements are actually compatible with each other within uncertainties.

The ${}^{22}{\rm Ne}(\alpha,\gamma){}^{26}{\rm Mg}$ reaction is also a key channel for the ${}^{22}{\rm Ne}$ destruction, directly competing with ${}^{22}{\rm Ne}(\alpha,{\rm n}){}^{25}{\rm Mg}$. The data for the ($\alpha, \gamma$) rate are analyzed in panel b) of Figure 4 in the same way as before \citep[with the exception of][as these authors didn't present an estimate for it]{jae01}. In the energy region of interest for stellar nucleosynthesis the rate suggested by \citet{lon12} is 30 $-$ 45\% higher than the NACRE results \citep{ang99} and the data of \citet{kae94}.

The ratio between the two destruction channels of $^{22}$Ne is crucial for estimating the number of neutrons available to the $s$-process. By adopting the most recent reaction rates discussed above the number of neutrons is expected to increase with respect to previous works. This increase will be small in comparison to NACRE or \citet{kae94}.

In the literature, the ${}^{18}{\rm O}(\alpha,n){}^{21}{\rm Ne}$ reaction is not considered as an effective neutron source for low mass stars, due to the prevailing ($\alpha,\gamma$) channel. Nevertheless, ($\alpha$,$n$) captures on $^{18}$O may play some role (as competitors) in the reaction network that controls the production of ${}^{19}{\rm F}$ and of ${}^{22}{\rm Ne}$ itself. The rate adopted in many stellar models is taken from the NACRE compilation \citep{ang99}, but is based on an unpublished measurement. A new experiment \citep{bes13}, performed at the Notre Dame Science Laboratory, revealed that a previously considered resonance (at $E_{\alpha}$ = 888 keV) was incorrectly attributed to ${}^{18}{\rm O}$ and had instead to be ascribed to ${}^{17}{\rm O}$. As a consequence, the estimate for the cross section of ${}^{18}{\rm O}(\alpha,n)$ is 20 $-$ 30\% lower than the NACRE value \citep{ang99}, resulting in an even smaller contribution to the neutron balance than previously assumed at the energies relevant for LMS evolution.

\begin{figure}
\centering
\includegraphics[scale=0.5]{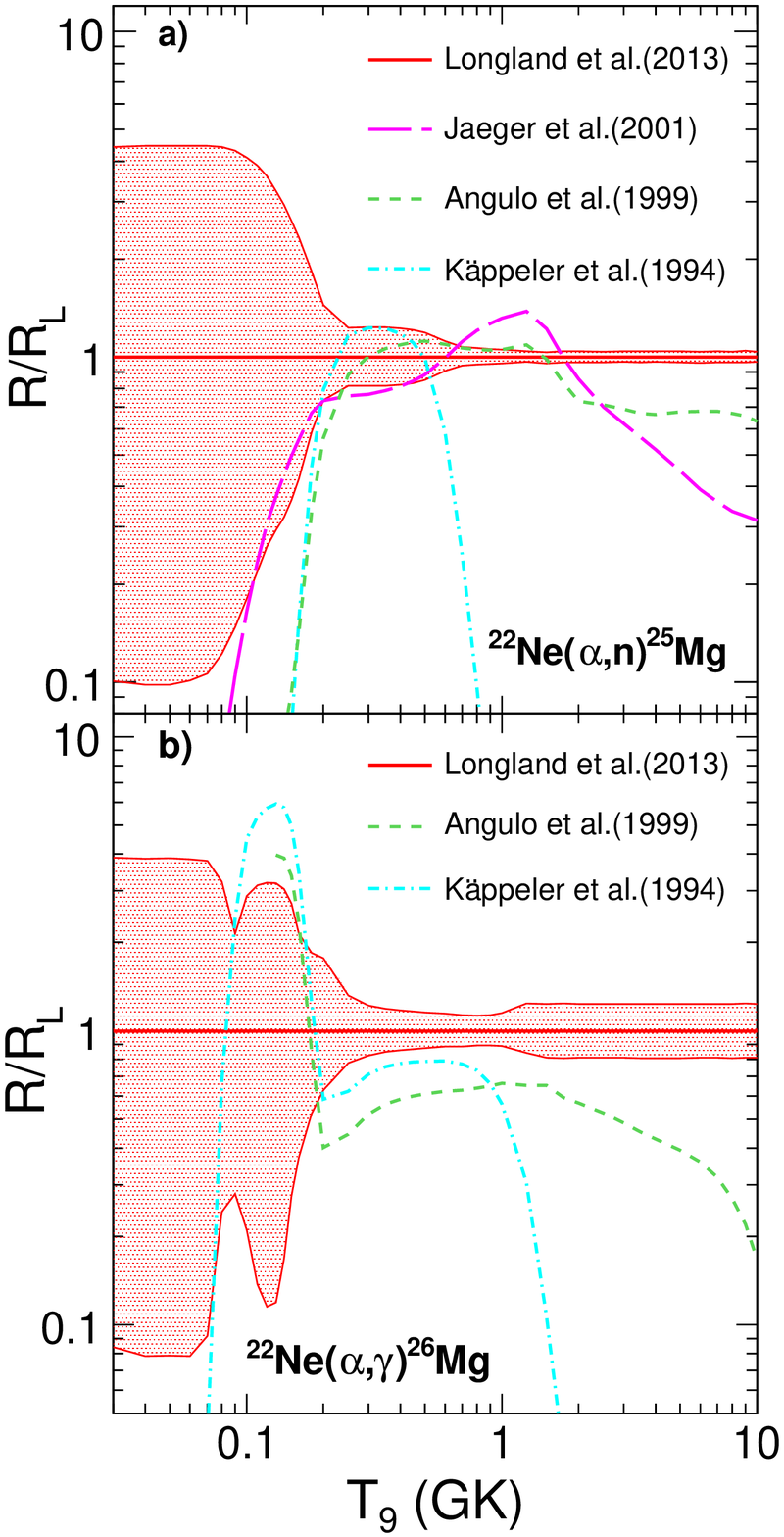}
\caption{\textit{Upper panel}: A comparison among the most recent reaction rates for the ${}^{22}{\rm Ne}(\alpha,{\rm n}){}^{25}{\rm Mg}$ reaction. The estimates by \citet{kae94} (dot-dashed, light-blue line), 
\citet{ang99} (short-dashed green) and \citet{jae01} (long-dashed magenta line) are plotted relative to the reference data by \citet{lon12} shown by the solid (red) lines. \textit{Lower panel}: the same ratios as above, but for the ${}^{22}{\rm Ne}(\alpha,\gamma){}^{26}{\rm Mg}$ reaction; notice that \citet{jae01} did not revise this rate. \newline
(A color version of this figure is available in the online journal.)} \label{four}
\end{figure}

The neutron capture cross sections adopted in our calculations are updated according to the KADoNiS database (2009 release) and using the subsequent literature. In particular we adopted the \citet{mas12} measurements for the magnesium isotopes (including the strong neutron poisons $^{25,\;26}$Mg), which were measured at the n$_{-}$TOF facility, at CERN.
For $^{74,\;76}$Ge and $^{75}$As the (n,$\gamma$) cross sections were taken from \citet{mar09}, while we adopted the new n$_{-}$TOF results for the Zr isotopes \citep{tag12}. For the osmium isotopes our references are the Maxwellian averaged cross sections from \citet{mos10}, plus the stellar enhancement factors by \citet{fuj10}.

Concerning weak interactions, our main source remains the compilation by \citet{tak87}. Subsequently published data were included for ${}^{79}{\rm Se}$ \citep{kla88}, ${}^{163}{\rm Dy}$ \citep{jun92}, ${}^{176}{\rm Lu}$ \citep{kla91,moh09}, ${}^{187}{\rm Re}$ \citep{bos96} and ${}^{207}{\rm Tl}$ \citep{oht02}.
Our choice for ${}^{176}{\rm Lu}$ requires a separate comment. Indeed the recent results by \citet{moh09} indicate a coupling of the isomeric state with the ground state 11 times faster than reported in \citet{hei08a}. As a consequence of the corresponding reduction in the branching factor $f_n$, \citet{moh09} found it difficult to reproduce the solar ${}^{176}{\rm Lu}/{}^{176}{\rm Hf}$ ratio  within the treatment of He shell flashes. As a general rule, however, we prefer to use experimental data when available, so that we continue to adopt the \citet{hei08a} estimates in this paper.

\section{Nucleosynthesis calculations for \textit{s}-processing in AGB stars.}

In order to compare the predictions from models adopting the two different ${}^{13}$C pockets discussed in section 2, we performed $s$-process nucleosynthesis calculations through our post-process code \textit{Nucleosynthesis of Elements With Transfer of Neutrons} (NEWTON),
which is an upgrade of the one adopted in \citet{bus99}. It includes a detailed network of more than 400 isotopes (from He to Bi) connected by $\alpha$-, $p$- and $n$-captures and weak interactions. The stellar evolutionary models for LMS in the AGB stages were taken from the FRANEC prescriptions \citep{str03}. 

When (for the sake of comparison) we deal with a pocket similar to those from \citet{gal98,bus01}, we need to adopt all the choices (e.g. the large number of thermal pulses) that were essential parts of that scenario, as summarized by \citet{kae11}. In the discussion of the new assumptions for the pocket, we can instead adopt a more modern view, which is now incorporated also in the FRANEC code \citep{cri09,cri11}. This view descends from the recent infrared analysis of AGB stars \citep{gua06, gua08, gua13}. In those works it was shown that efficient mass loss prevents the  AGB luminosities to attain values larger than about 10$^4~L_{\odot}$, thus implying a lower number of pulses than in previous models. Revisions of the opacities now also guarantee a larger efficiency of TDU episodes. As a consequence, the total amount of processed matter is about the same as before, but the present models now reproduce theoretically the Luminosity Function (LF) of C stars. There might be actually a slight overestimate of mass loss rates in the new cases as computed by FRANEC (Guandalini, private communication), so that we perform $s$-process calculations for 3 $-$ 4 pulses more than reported in the on-line repository of the FRANEC data \citep[\textit{http://fruity.oa-teramo.inaf.it}, ][]{cri11}, using the parameters of the last pulse computed. With the above approach, a ``new" 1.5 \msb case experiences 8 pulses less than in the choices by \citet{bus01}.

In our LMS models, the ${}^{13}{\rm C}(\alpha,{\rm n}){}^{16}{\rm O}$ reaction is activated in radiative conditions (at about 8 keV), during the periods between two subsequent convective instabilities. Except for rare situations occurring at very low metallicity, ${}^{13}$C is consumed locally, before the onset of the subsequent pulse. The two stellar evolution scenarios we discuss here for $s$-processing differ only for: i) the temperature of thermal pulses, which is slightly lower in the new cases; and ii) the number of pulses and
the corresponding TDU efficiencies, as explained above. The cumulative effects induced by these differences on the final yields are very small, so that any change emerging in the final production factors can be safely interpreted as due primarily to the different extensions of the pocket in the two cases. 

As mentioned, for the old cases, we adopt a pocket of 10$^{-3}$ \ms. As a consequence of p-captures in those layers, the resulting integrated amount of ${}^{13}$C available for producing neutrons is about $3.45\times10^{-6}$ \ms. With this \ctb abundance the yields of LMS mimic the solar distribution at about [Fe/H] $\simeq -0.5$; the distributions one can find in this way (see later Figure 5) are very similar to that by \citet{bis10} and the accuracy is roughly the same. We notice that this is not much different from the recent results by \citet{bis14}, who indeed continue to depict a scenario for $s$-processing incapable of explaining the light $s$-elements and incapable of reproducing the abundances in open clusters. Their models are therefore, in the present context, not much different that the older ones we quoted. We shall refer to the calculations performed with the above assumptions as to ``Case A". When adopting the larger \ct-reservoir, we base our analysis on the previous calculations by \citet{mai12}. There it was shown that the bulk of the $s$-process abundances now seen in the Galaxy was produced by AGB stars with metallicity with a small range ([Fe/H] $= -0.15$ to $-0.12$); the value [Fe/H] $= -0.15$ is adopted here. In this case the choice for the pocket is that presented in Figure 2 and extends almost to the whole He-rich layers. We shall refer to the calculations performed with these assumptions as to ``Case B".

For both choices of the pocket, we computed the outcomes of $s$-process nucleosynthesis in model stars of 1.5, 2.0 and 3.0 \ms. In the calculations we used the compilation of solar abundances from \citet{lod}, implying  $Z/X \simeq 0.0215$. A subset of results (relative to the extreme cases of 1.5 and 3.0 \msb models) are shown in Figures 5 and 6, for Case A and Case B, respectively. These data represent the production factors ${O}^{i}={X}^{i}/{X}^{i}_{\odot}$ in the He-rich region at the last computed thermal pulse, for the nuclei in the atomic mass range $70 \lesssim {A} \lesssim 210$, with respect to the initial abundances. For comparison, the production factors are normalized to the mean value of those $s$-only nuclei that are not affected by branchings (in Figures 5 and 6 we use ${\overline{X}}_{\rm sowb}$ to refer to ``\textbf{s}-\textbf{o}nly nuclei \textbf{w}ithout \textbf{b}ranchings''). This means that ratios of unity indicate the exact level necessary to fit the solar system distribution of $s$-only nuclei. The choices discussed above also imply that a flat distribution for all $s$-only nuclei corresponds to a good average model for the main $s$ component. To guide the eye, a tolerance region of $\pm$ 10\% is indicated by the red dashed lines in Figures 5 and 6.

The enhancement factors with respect to the initial composition found at the last computed pulse of each model was then weighted using the value of the Salpeter's Initial Mass Function (IMF) corresponding to the stellar mass and summed with those from the other masses. The result was then normalized, again by dividing for the average over-production of $s$-only nuclei unaffected by reaction branchings, in order to construct the (weighted and normalized) average production from the whole mass range. The results of the averaging procedure are shown in Figure 7 for Case A and Case B, respectively. They are similar to the results shown in Figures 5 and 6 for 1.5 \msb models; this is obviously due to the weighting operation, as the IMF favors lower masses. We therefore confirm the common assumption that models for this stellar mass offer an average rather typical conditions for producing the main component.

As already mentioned, the stellar models at the base of Case A and Case B are very similar; in particular, the temperatures of the radiative layers where most of the neutron flux is released are essentially the same. As also the nuclear parameters adopted are the same, any difference is due primarily to the extension of the  ${}^{13}$C reservoir.  

\begin{figure}[t!]
\centering
{\includegraphics[scale=0.4]{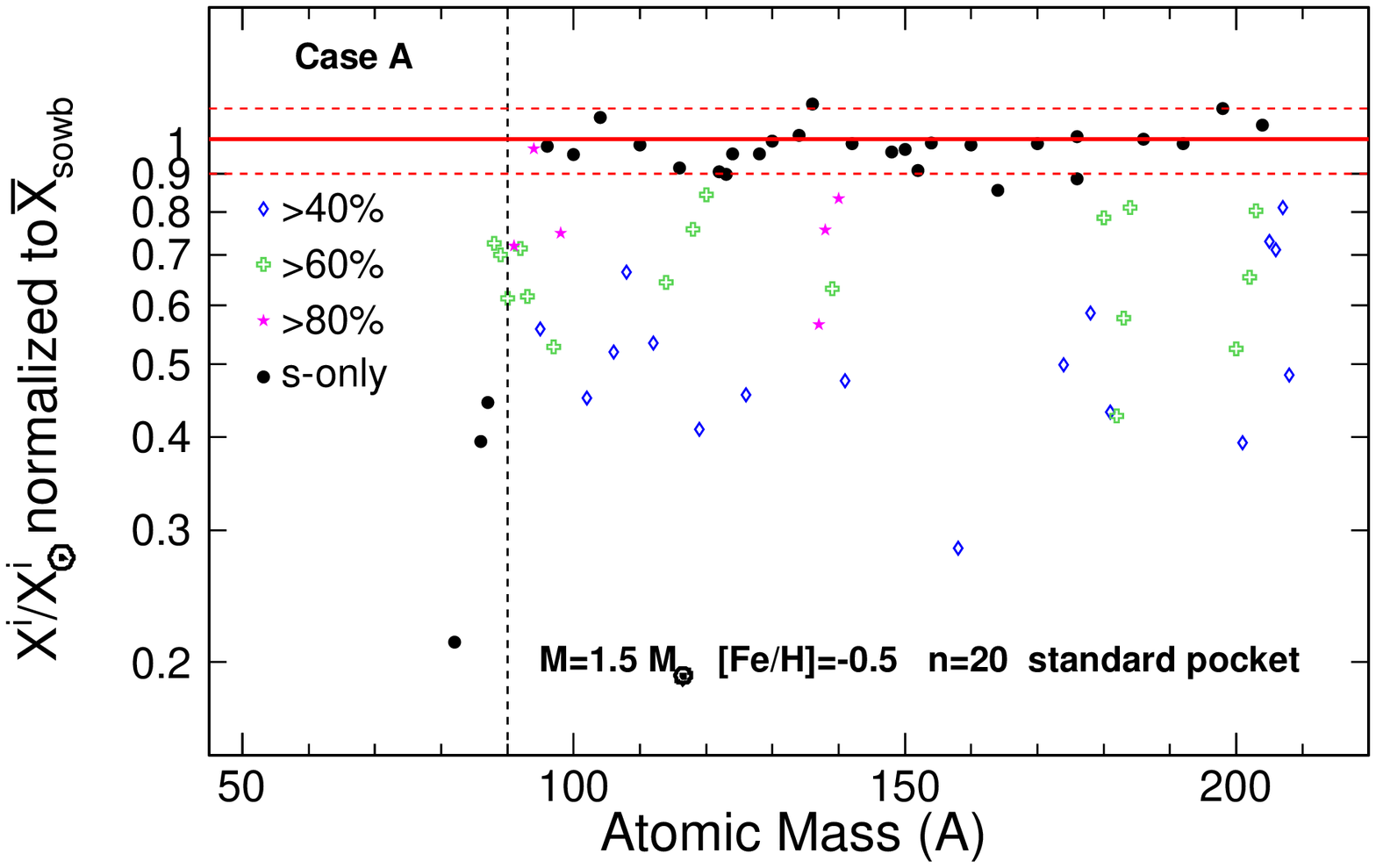}
\includegraphics[scale=0.4]{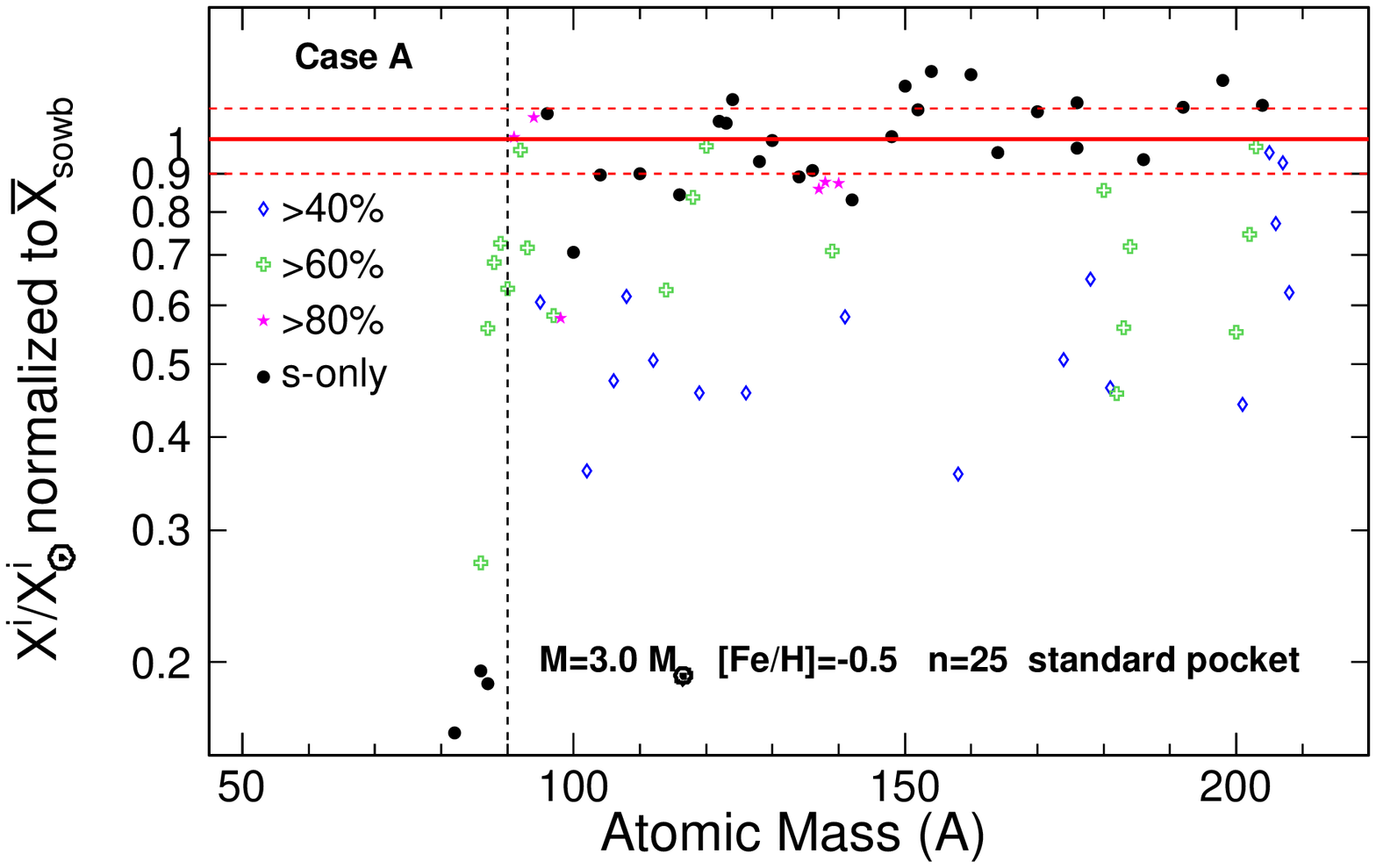}}
\caption{The $s$-element production factors with respect to initial abundances, in the He-intershell layers, from stellar models of 1.5 and 3.0 \msb after the last computed thermal pulse, using the smaller \ctb pocket mentioned in the text. A large number of pulses (indicated by ``n'' in the figures) is assumed, as in the original works (see discussion in the text). As the number of available neutrons is small, the efficiency of the process becomes suitable to fit the solar distribution only for moderately low metallicities ([Fe/H] $\simeq$ $-0.5$). In the figure, only nuclei with expected $s$-process contributions larger than 40\% are shown.\newline
(A color version of this figure is available in the online journal.)}\label{five}
\end{figure}

In Figures from 5 to 7 we used different symbols (and different colors in the electronic version) to distinguish the isotopes according to the so-far expected percentage production by slow neutron captures (see the legends in the Figures). A vertical dashed line was drawn to indicate the starting point for the main component, conventionally defined at ${\rm A} = 90$ \citep{kae11}. 
 
\begin{figure}[t!]
\centering
{\includegraphics[scale=0.4]{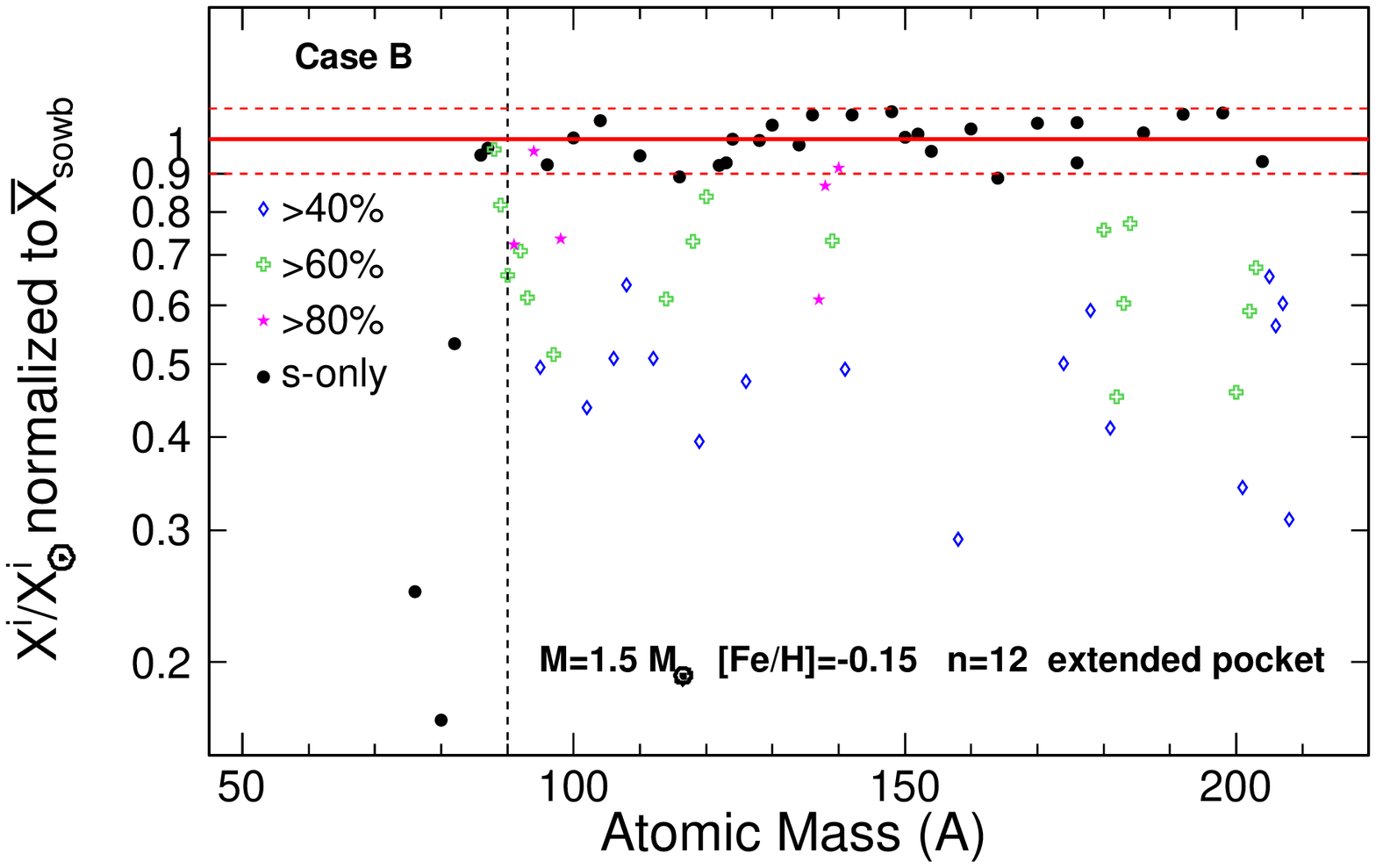}
\includegraphics[scale=0.4]{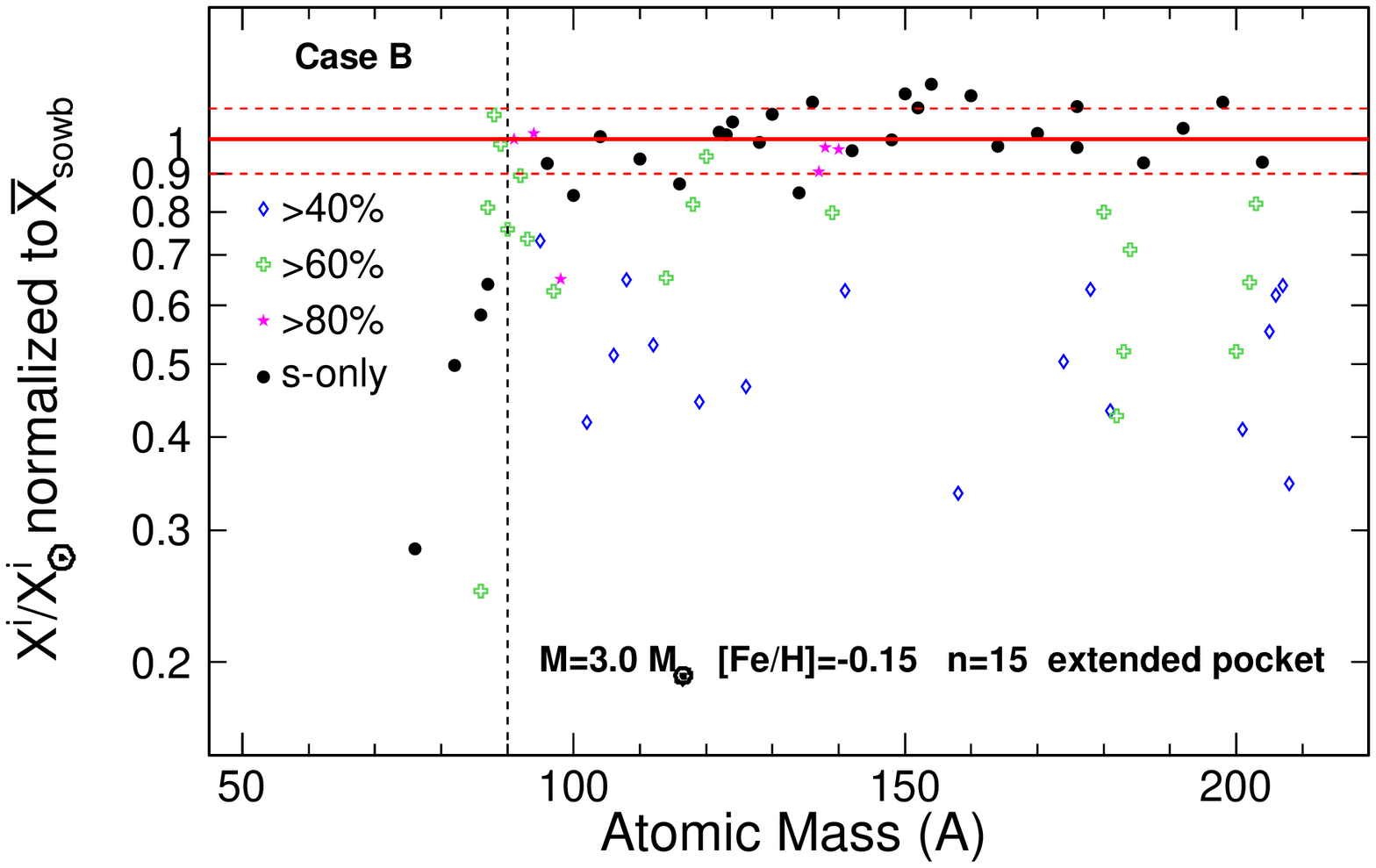}}
\caption{Same as in Figure 5, but using the extended \ctb reservoir described in the text. Here, for obtaining a nearly-solar distribution of $s$ elements, a smaller number of cycles and a higher metallicity ([Fe/H] $\simeq$ $-0.15$) can be adopted. The stellar masses are the same as in Figure 5. \newline
(A color version of this figure is available in the online journal.)}\label{six}
\end{figure}

\section{Reproducing the main component}

\begin{figure}[t!]
\centering
{\includegraphics[scale=0.4]{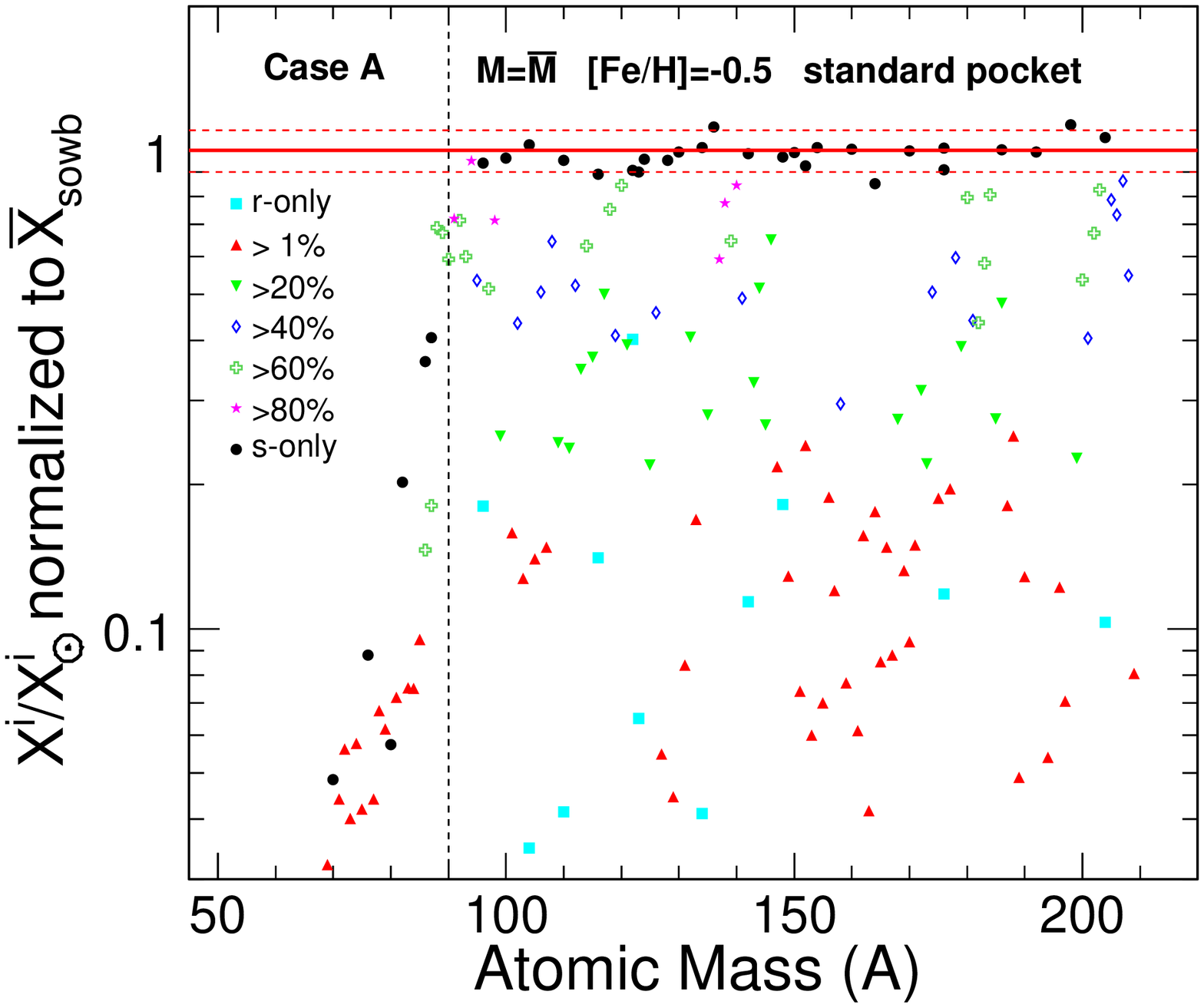}
\includegraphics[scale=0.4]{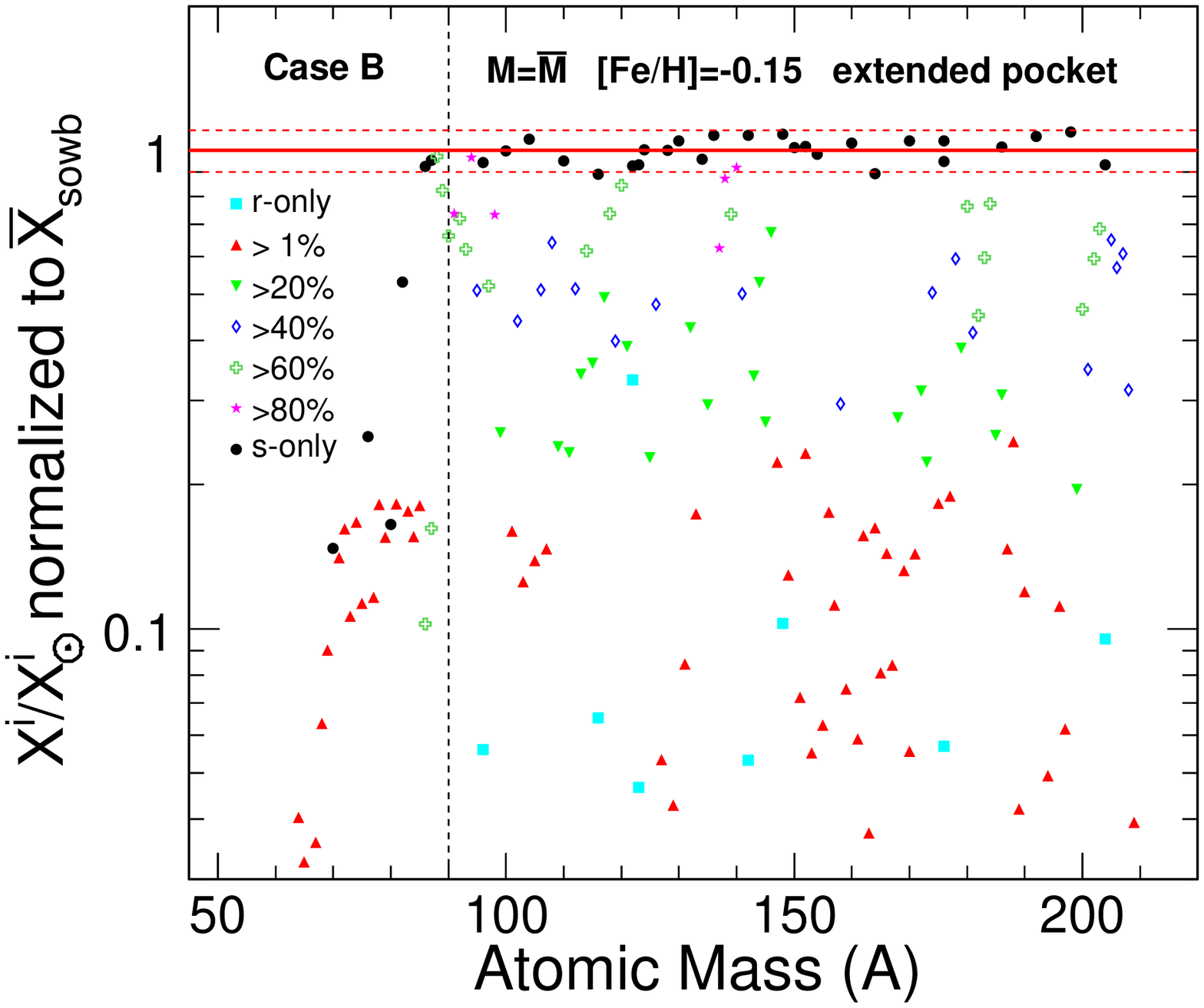}}
\caption{Left panel: the distribution of normalized production factors from the models of Figure 5, once averaged over the mass interval from 1.5 to 3 \ms, using the Salpeter's IMF for weighting the data. 
Right Panel: the same kind of distribution, obtained from a similar average of the new models shown in Figure 6.\newline
(A color version of this figure is available in the online journal.)}\label{seven}
\end{figure}

\begin{figure}[t!]
\centering
\includegraphics[scale=0.45]{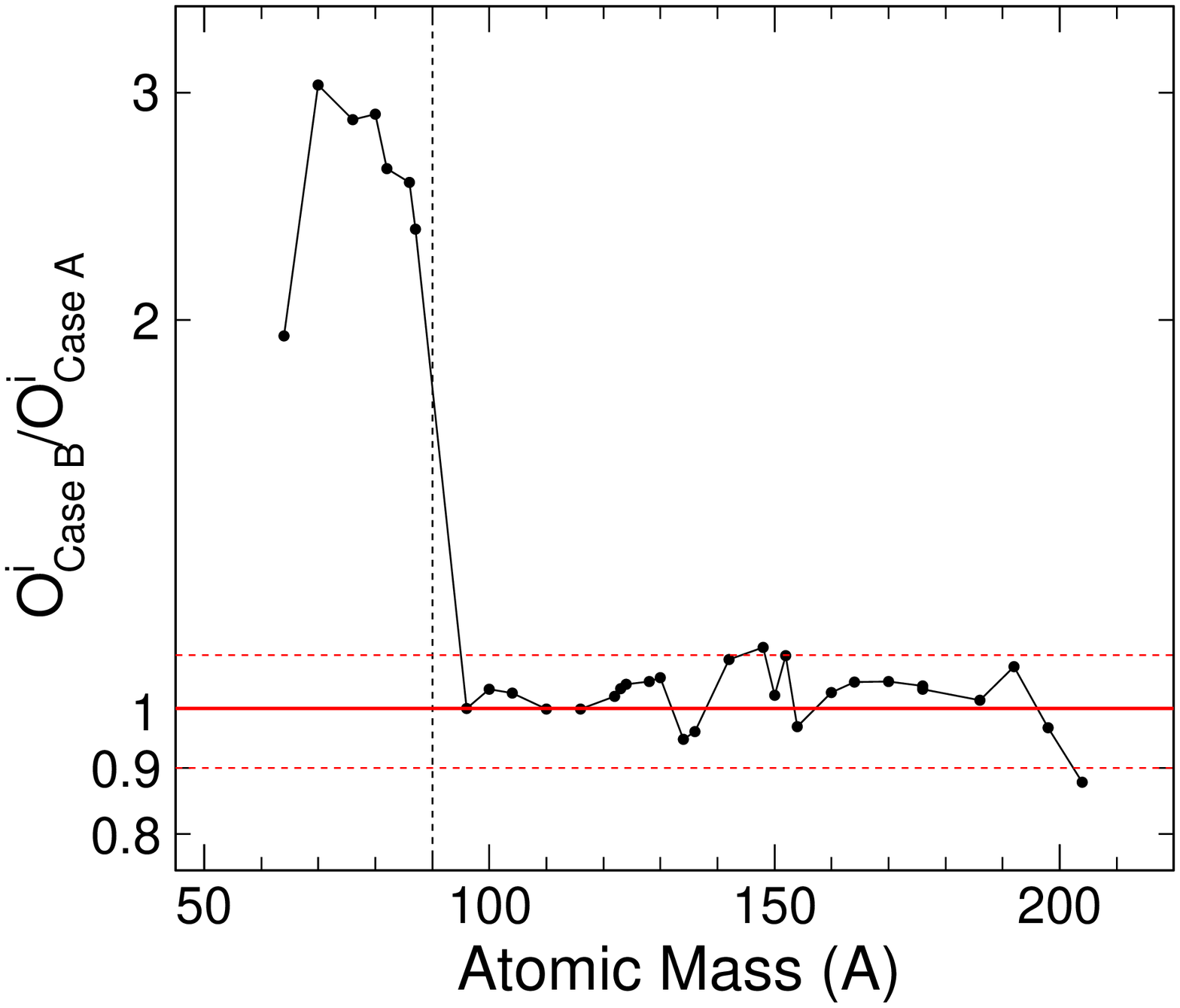}
\caption{The ratio of the normalized production factors for the $s$-only nuclei plotted in Figure 7 for Cases B and A. While both models are in fair agreement in the mass range $A \gtrsim 90$, the case with the more extended ${}^{13}$C pocket yields large contributions to nuclei below $A \simeq$ 90, thus providing expectations for the weak $s$-process component in massive stars.\newline
(A color version of this figure is available in the online journal.)}\label{eight}
\end{figure}

A direct comparison between the panels of Figure 7 immediately highlights that the two distributions
differ significantly only outside the atomic mass limits of the main component (i.e. for A $\lesssim$ 90 and A $\gtrsim$ 204). This result is in itself remarkable. It means that, by using a large \ctb pocket, already suitable to account for the trend of $s$-process abundances in the Galactic disk, up to the youngest clusters, it is also possible to reproduce quite well the solar distribution and in the mean time to incorporate the upgrades on AGB modeling necessary to account for the LF of AGB stars. 

The differences between the two sets of results are summarized in the ratios of the production factors for the $s$-only nuclei, as plotted in Figure 8. Again, the vertical dashed line represents the A $=$ 90 starting point of the main component, while the red horizontal dashed lines identify a 10\% fiducial interval. This is the typical level of the experimental uncertainties on the points to be fitted, when one considers both abundances and nuclear parameters \citep[see e.g.][and references therein]{kae11}. As one can notice, for A $\geqslant$ 90 the situation is virtually the same and the ratios differ from unity only by a few percent. An apparent exception is seen at ${}^{204}$Pb, but it only derives from the fact that the ratio emphasizes small differences pointing to opposite sides of the average level.

On the contrary, for A $\leq$ 90 Case B feeds more efficiently the lighter (or weak-$s$) elements, resulting in two to three times higher production factors.  This point deserves some comments. As shown by \citet{tra04}, Case A implies that AGB nucleosynthesis becomes insufficient to explain the Galactic enrichment of $s$-process nuclei below A $\simeq$ 100 $-$ 120 and an unknown process, called solar $LEPP$ (Light Element Primary Production), must be invoked. Consequently, predictions for the weak $s$-process contributed by massive stars cannot be derived by our computations. For Case B, instead, integrations from unknown processes 
are not needed \citep{mai12}, so that Figure 7 can be used to predict the role played by massive stars 
in the synthesis of light $s$-nuclei. This role depends crucially on still poorly known nuclear parameters, especially the $^{12}$C($\alpha$,$\gamma$)$^{16}$O rate; hence, firm constraints from AGB stars might serve as guidelines for the expectations on such parameters. 

We warn that the predictions for the weak component here derived depend on the mass of the \ct-pocket, for which only an exemplifying average extension was chosen from \citet{mai12}. More precise indications should be derived with a dedicated analysis.   

Among the relevant properties of the abundance distribution for Case B we underline the contributions to the the $s$-only isotopes of strontium, ${}^{86,\;87}$Sr. Their production factors are increased by a factor of almost 3 with respect to Case A and stay very close to the reference line of $s$-only nuclei produced by the main component. In our new scenario, therefore, the matching point between the $s$-element production from massive and AGB stars would require to be moved downward, at the $^{85}$Kr branching; ${}^{86,\;87}$Sr would in this case become full members of the main component.
A further general property is that, due to the large neutron exposure, all nuclei near magic numbers are fed very efficiently. Hence, the contribution by the $s$-process main component to $^{88}$Sr, $^{89}$Y and $^{94}$Zr becomes close to unity (about 95\%, 83\% and 97\%, respectively). The same effect is seen near the $N=82$ magic number, where $^{138}$Ba and $^{140}$Ce are now produced at the 88\% and 92\% level, respectively, by the $s$ process. ${}^{208}$Pb represents a special case, although it is not an $s$-only nucleus and therefore is not crucial for the solar-system $s$-element distribution. Its $s$-process contribution does not come only from the main component, as discussed by \citet{gal98}. These authors also showed how its production from the so-called ``strong'' $s$-process component could be provided by low metallicity AGB stars. In view of this crucial role of the metallicity dependence and of the fact that Cases A and B have a different reference metallicity, the two production factors shown in Figure 7 cannot be compared directly, as the parent stars would have very different roles in the chemical evolution of the Galaxy for lead. In this case the contributions of AGB stars to the solar abundance can only be derived by Galactic chemical evolution calculations.  

We conclude this section by summarizing the main results found here: i) the solar distribution of $s$-process isotopes from Zr to Pb is mimicked well by nucleosynthesis calculations made for LMS undergoing thermal pulses in AGB phases. ii) The metallicity at which the yields from AGB stars best approach the distribution in the Sun increases with the extension of the \ctb pocket. iii) If the pocket is sufficently large, the dominant metallicity is in the range typical of the most common Galactic disk stars and the yields dominate the Galactic enrichment integrating the weak $s$-process in massive stars. iv) Only for small extensions of the \ctb pocket another independent nuclear process (the solar $LEPP$) is required to complement the AGB production in the mass range 100 $\leq$ A $\leq$ 120. v) Constraints from stellar luminosities and from recent results on mixing at the envelope border \citep{pie13} play in favor of the scenario with the large pocket inferred by \citet{mai12}, which combines a limited number of pulse-interpulse cycles with a high processing efficiency and does not require any $LEPP$ contribution. vi) This new scenario foresees important contributions by AGB stars to the light $s$-process nuclei of the weak $s$ component.

\section{Discussion and Conclusions.}
In this paper we re-analyzed nucleosynthesis models for slow neutron captures in AGB stars, after new observational as well as theoretical information shed doubts on the previous scenario for the formation 
of the \ctb neutron source and for its actual extension. In our work we have argued that, even in 
presence of persisting uncertainties concerning the dynamical mechanisms promoting proton penetration 
into the He-rich layers at the convective border, stellar physics offers other, perhaps more secure, 
ways of generating transport phenomena suitable for forming a \ctb reservoir and then inducing neutron-capture nucleosynthesis. In particular, we have suggested that a fruitful line of research may be that of describing, through a quantitative MHD treatment, the development of toroidal magnetic fields, induced by stellar dynamos, in the radiative He-rich layers below the convective envelope. The above scheme for the creation of a \ct-rich layer foresees that the partially mixed zones extend down to very deep regions, essentially involving most of the He-rich layers of the AGB star, due to the formation of buoyant magnetic structures close to the outer border of the degenerate C-O core.

We have also underlined that any attempt at upgrading our present understanding of $s$-processing in low-mass AGB stars must take into account the fact that the infrared LFs of these last agree with stellar model predictions only if the magnitudes remain moderate ($M_{bol} \lesssim - 5$) and hence the number of pulses undergone by the star is smaller than previously assumed \citep{gua13}. The above considerations imply that $s$-processing in AGB stars is built in a way rather different than imagined so far, namely through a smaller number of pulse-interpulse cycles, each however experiencing a more efficient nucleosynthesis episode. As these required changes are also necessary to explain the increasing abundances of $s$-elements in the Galactic disk \citep{mai12}, they seem to become mandatory. Also, they cannot be mimicked by increasing the abundance of \ctb in a small pocket: the concentrations in mass of \ctb and of $^{14}$N in the reservoir formed are fixed by H-burning rates and cannot be varied freely (as is instead often done), without violating basic physical laws.

We have then performed a comparison of the results achievable for reproducing the solar main component in two cases: i) the scenario most commonly used in the last 20 years for dealing with s-processing, based on a small extension of the \ctb reservoir (that we called Case A); and ii) the newly suggested one, with \ct-rich layers reaching down to deep regions of the He-rich zone (referred to as Case B). The comparison has been performed by computing  n-captures in stellar models of 1.5, 2.0 and 3.0 \msb and by averaging the $s$-process results after weighting by the IMF of Salpeter. 

The main result is that, if the metallicity for the two cases is chosen suitably, both provide a distribution of production factors mimicking the main $s$-process component. Due to the different neutron capture efficiency resulting from the different extension of the \ctb pocket, the number of pulses differs in the two cases, much like the metallicity does: [Fe/H] $\simeq - 0.5$ and n $\gtrapprox$ 20 pulses for the old scenario, [Fe/H] $\simeq - 0.15$ and n $\lessapprox$ 15 pulses for the new one.

Moreover, the main aim of the above test was to look for an answer to the question posed in the title: can we distinguish, from comparisons with solar abundances, which scenario has to be preferred? In general, if one sticks to the results from a stellar generation at a suitably chosen metallicity then a decision is not possible, as the quality of the fits to the solar abundances of $s$-only nuclei shown in Figure 7 is essentially identical.

\begin{figure}[t!!]
\centering
\includegraphics[scale=0.45]{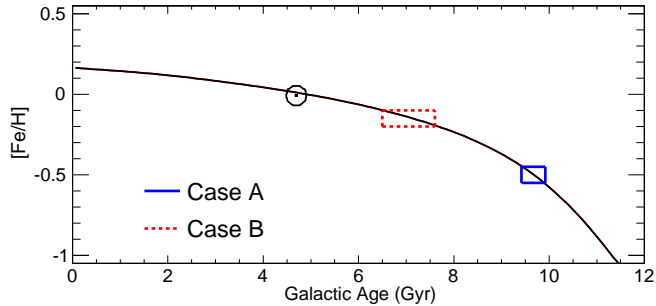}
\caption{The Age-Metallicity Relation (AMR) in the Galactic disk as derived by \citet{mai12}. The two boxes roughly indicate the metallicity and age intervals where the main component is best fitted by AGB star nucleosynthesis, in the cases (A and B) discussed in the text. The symbol of the Sun is also shown in the figure. \newline
(A color version of this figure is available in the online journal.)}\label{nine}
\end{figure}

However, a closer look reveals remarkable differences in the predictions of the two cases for the nucleosynthesis of $s$-nuclei in the Galaxy. This is already evident from Figure 8, if one considers light nuclei outside the limits of the main component; but is true also for heavier isotopes, when one derives the consequences of the calculations for the chemical evolution of the Galaxy \citep{tra04,mai11,mai12}. Both issues are actually strictly connected as outlined in the following. Let us show how.

The reason for the different predictions from Case A and Case B at the lower mass end of the distribution (requiring or not a solar $LEPP$ process) is rather simple. It can be illustrated with the help of the Age-Metallicity Relation (AMR), which is reproduced in Figure 9 from the results by \citet{mai12}. The two boxes represent the metallicity intervals over which the main component is best fitted in our Case A 
and Case B. For Case A the AMR is sampled over a short time interval, at epochs old enough that it is still far from the conditions prevailing over most of the Galactic disk duration. The total number of stars in that short interval is therefore relatively small and the effects on Galactic abundances will not be dominant.  Most AGB stars will be born later, when the abundance of Fe is higher. Due to the small pocket, the number of neutrons per iron seed in them will be so small that their yields will be almost irrelevant in the global inventory of the Galaxy. As, with low neutron exposures, they feed mainly light $s$-process nuclei, these last will be insufficiently produced, hence the requirement of a $LEPP$ integration. On the contrary for Case B the reference metallicity range, due to the large pocket, is shifted upward, to conditions typical of the main Galactic disk population, lasting for several Gyr. In this case the AGB stars shown before to mimic the main component will be the dominant ones, sufficient in number and effectiveness for taking care of the Galactic enrichment, so that no extra process is required. These are examples of a more general trend. Essentially, by choosing adequately the metallicity and the \ctb pocket extension, one can obtain production factors mimicking the solar distribution in generations of AGB stars for any choice of the \ctb reservoir. However, if we want that the chosen generation can process enough Galactic material to be really dominant in the chemical evolution of the Galaxy and in controlling solar abundances, then we must choose an effective average metallicity typical of the thin Galactic disk, where the abundance evolution is low,  long time scales are involved and the number of AGBs contributing becomes huge. In that case, the abundance of iron is high and to have a sufficient number of neutrons per iron seed the \ctb pocket must be quite extended in mass. This favors Case B. Obviously, Case A cannot be excluded on these grounds, but it would need a $LEPP$ contribution. For Case A, this means that searching for $average$ Galactic conditions where the solar distribution is reproduced is not really meaningful.

The above discussion gives us an opportunity to identify crucial tests that should be made, from which a conclusive judgement can be derived on the real extension of the \ctb pocket (hence also on its origin). 
We list below six such tests that are, in our opinion, especially suitable to provide a final answer.
\begin{itemize}
\item
{Compute models using the Case A choice for the pocket, but with a limited number of pulses (thus reaching a luminosity compatible with present-day LFs), verifying whether a compromise can be found that fits the solar data without violating the prescriptions on C-star magnitudes. We believe this should be actually very difficult, given the shortage of neutrons; but this is in any case a crucial test to be performed quantitatively.}
\item
{Compute Galactic chemical evolution models including at least Sr, Ba and Pb isotopes with the two scenarios and compare them with the observations (which are unfortunately limited for Pb abundances). Very young stellar clusters (absent in previous such studies by \citep{rai99,tra01,tra04} should be included. We expect that the models of Case A will not reproduce the observations, while those of Case B will; but again this has to be demonstrated in detail.}
\item
{Verify whether, with an extended \ctb pocket, one can reproduce the $s$/C ratios of post-AGB stars, an achievement that proved impossible for the  models of Case A \citep{per12}.}
\item
{Detailed, quantitative models (based on MHD calculations or on other processes capable of forcing the formation of a \ctb pocket) should be developed to see what kind of mixing can be realistically expected.}
\item
{The abundance pattern shown by presolar materials recovered in pristine meteorites should be compared with the predictions of the two scenarios, looking for more detailed constraints possibly coming from the isotopic admixtures measured in presolar grains.}
\item
{When the chemical evolution of the Galaxy is computed, models of Case A were shown to require, for explaining the solar system abundances of $s$-elements up to A $\simeq$ 120, the contribution of the unknown solar $LEPP$ process \citep{tra04}. This is not necessarily coincident with the process required at low metallicity, see e.g. \citet{mon07}.
From the tests made on crucial elements by \citet{mai12} we know this is not needed by the new models of Case B. Now a critical point is: can the approach of Case A, plus a unique choice for the LEPP contribution, explain the increased abundances of $s$ elements in young Galactic stellar systems? An answer can come from fixing the LEPP contributions from solar constraints, then verifying if this is sufficient for explaining the increased abundances in young clusters. Again we predict that this procedure should fail and the results by \cite{bis14} seem to point in that direction. However they do not consider the open cluster problem directly, so that a dedicated calculation must be done before a final decision is taken.}
\end{itemize}

The information we can get from performing the above tests would be decisive. Should the new models,
with an extended \ctb pocket and a limited number of pulses, prevail (as it may seem probable now, given 
the larger number of constraints they appear to match) then some general conclusions on $s$-processing 
should be revised. In particular: i) the main component should be considered as including $^{86,\;87}$Sr completely; ii) the expectations for $^{208}$Pb in low metallicity stars would be different and probably 
less extreme; iii) the $s$-process contribution to nuclei like $^{88}$Sr, $^{89}$Y, $^{94}$Zr, $^{138}$Ba and $^{140}$Ce should be revised upward and accepted to reach 90 $-$ 100\% of their abundance; iv) in view of the expected new measurements for the rates of the $^{12}$C($\alpha,\gamma$)$^{16}$O and of the $^{12}$C+$^{12}$C fusion reactions, new determinations of $s$-processing in massive stars should verify the new suggestions from AGB models for nuclei below A$\sim$ 90.

{\bf Acknowledgments.}

O.T. thanks $INFN$ \textit{Istituto Nazionale di Fisica Nucleare}, section of Perugia, for financial support and Marco La Cognata for constructive discussions on the ${}^{13}{\rm C}(\alpha,{\rm n}){}^{16}{\rm O}$ reaction rate. S.P. acknowledges support from the Italian MIUR through the excellent found for Nuclear Astrophysics ``LNS Astrofisica Nucleare - fondi premiali''. The authors are grateful to the referee for a very careful and constructive report, that greatly helped in clarifying many relevant issues.

\end{document}